\documentclass[a4paper,11pt]{article}
\pdfoutput=1

\usepackage{jcappub}
\usepackage[T1]{fontenc}

\usepackage{amsmath, amssymb}
\usepackage{graphicx}
\graphicspath{{./figs/}}
\usepackage{caption, booktabs, subcaption}

\usepackage{hyperref}
\usepackage{cleveref}

\usepackage{multirow}
\usepackage{array}
\usepackage{siunitx}

\usepackage{ulem}
\usepackage{color}
\definecolor{orange(colorwheel)}{rgb}{1.0, 0.5, 0.0}

\title{\boldmath Hamilton-Jacobi Approach to Inflationary Scenarios through Extended Entropies: An Observational Perspective}

\author[a,1]{H. R. M. Zarandi,\note{Corresponding author.}}
\author[b]{E. Ebrahimi}
\author[c]{Y. Toda}

\affiliation[a]{Department of Physics, College of Science, Shahid Bahonar University, PO Box 76175, Kerman, Iran}
\affiliation[b]{Department of Physics, College of Science, Shiraz University, Shiraz 71454, Iran}
\affiliation[c]{Kochi University of Technology, Department of Data and Innovation, Tosayamada 782-8502, Japan}

\emailAdd{hrmz0072@gmail.com}
\emailAdd{es.ebrahimi@shirazu.ac.ir}
\emailAdd{y-toda@particle.sci.hokudai.ac.jp}

\abstract{
The slow-roll inflation paradigm can be systematically generalized within the framework of non-standard entropy formalisms, giving rise to a broad class of inflationary models that deviate from the conventional Bekenstein--Hawking case. We adopt a pragmatic observational strategy, employing the Hamilton--Jacobi formalism to establish a direct link between the inflationary potential, the generalized entropy function, and the resulting cosmological observables. In this approach we introduce a novel non-linear parametrization of the Hubble parameter, yielding sensible results, including consistency with recent observational data and new estimates of the cosmological parameters of the generalized entropy framework: the Tsallis parameter $\delta\simeq1.1-1.2$, the R\'enyi parameter $\alpha\sim\mathcal{O}(10^{-14})$, and the Kaniadakis statistics parameter $K\sim\mathcal{O}(10^{-17})$. Our analysis proceeds in two regimes: first, by constraining models directly with the primary inflationary parameters including the scalar spectral index ($n_s$) and the tensor-to-scalar ratio ($r$); second, by exploring the impact of the observational uncertainty on the upper bound of $r$ ($\sigma_r$), which we vary to assess its influence on parameter estimation. This dual approach yields complementary posterior distributions that restrict the viable parameter space of entropy-based inflationary models. We further highlight the implications of the Hamilton--Jacobi method for the dynamics of the inflationary epoch, the reheating process, and, as a secondary objective, the subsequent evolution of cosmic structure in the late universe.
}

\begin{document}
\maketitle
\flushbottom

\section{Introduction}
\label{sec:intro}

The inflationary paradigm, first proposed by Starobinsky \cite{Starobinsky:1980te} through the inclusion of an $R^2$ term in the Einstein-Hilbert action. This seminal work, motivated by semi-classical quantum gravitational effects, was soon followed by Sato's scenario \cite{Sato:1980yn,Sato:1981ds} of rapid primordial expansion transitioning to a hot universe. Guth \cite{Guth:1980zm} subsequently formalized the idea of inflation as a solution to famous and key problems in Big Bang cosmology, the horizon, flatness and monopole problems through the so-called ``old inflation'' \cite{kazanas1980dynamics}. It is known that the old inflation suffers from phase transition from false vacuum to the true vacuum \cite{Guth:1980zm}. Four decades later a presented model of inflation \cite{Albrecht:1982wi,Linde:1981mu} established as one of the two fundamental pillars of modern cosmology, alongside late-time acceleration \cite{SupernovaCosmologyProject:1998vns}. Although inflation was accepted in the literature but its fundamental origins really seems enigmatic. The dominant paradigm invokes slow-roll dynamics of a scalar field component \cite{Linde:1983gd}. Intriguing connections between inflationary physics and the dark energy (the component in charge of the late cosmic acceleration) models have gained a notable attention \cite{Odintsov:2025jfq,Kaneta:2025dcs,Ghosh:2024jvs,Li:2023pfb,Keskin:2018awq}.

In this paper, we introduce the Hamilton--Jacobi formalism \cite{Salopek:1990jq,Langlois:1994ec,Rigopoulos:2003ak,Videla:2016ypa,Artigas:2025nbm} as an efficient method for analyzing inflationary scenarios.  
In this approach, the scalar field $\phi$ itself is treated as the time variable.  
This assumption is well justified during the early stage of the slow-roll regime, where key observables such as the scalar spectral index $n_s$ and the tensor-to-scalar ratio $r$ are determined.
A consequence of this formalism is that the Hubble parameter is specified as a function of the scalar field, $H(\phi)$, rather than specifying the potential $V(\phi)$ directly \cite{lidsey1997reconstructing}.  
This enables a more general and model-independent analysis, without the need to assume a specific functional form of the potential.  
In this work, we use the Hamilton--Jacobi approach to identify the parameter regions consistent with current observational constraints.

Using this model-independent formalism, we further investigate how entropy extensions affect the evolution of the inflationary epoch.

The profound connection between black hole thermodynamics and gravitational theory has improved our understanding of quantum gravity through two important developments: the AdS/CFT correspondence \cite{Maldacena:1997re,Aharony:1999ti} and the gravity-thermodynamics conjecture \cite{Padmanabhan:2003gd}. These frameworks provide concrete realizations of the holographic principle \cite{Susskind:1994vu}. Applying the thermodynamic-gravity conjecture to spacetime horizons, Einstein's field equations emerge as thermodynamic identities \cite{PhysRevLett.75.1260,Padmanabhan:2002sha,Eling:2006aw,Akbar:2006er,Padmanabhan:2007en,Padmanabhan:2009vy}. This approach has been extended to cosmology, where Friedmann equations can be derived by applying thermodynamic laws to the Universe's apparent horizon \cite{GarciadeAndrade:2000vz,Karami:2010zz,Moradpour:2016tuw,Sanchez:2022xfh}. While the original formulation employs Bekenstein-Hawking (BH) entropy $S_{BH} = A/A_0$ (that $A$ and $A_0$ are the horizon and Planck areas respectively), recent advances have explored generalized entropy forms motivated by non-extensive thermodynamics \cite{Verlinde:2000wg,Wang:2001bf,Cai:2005ra,Akbar:2006kj,Cai:2010hk}.

Among these generalizations, Barrow entropy \cite{Barrow:2020tzx} represents a particularly intriguing modification, expressed as $S = (A/A_0)^{1+\Delta/2}$ where $0 \leq \Delta \leq 1$ encodes quantum gravitational corrections. Though originally proposed for black holes, this entropy form has found significant cosmological applications, leading to modified Friedmann equations that predict novel phenomenology \cite{Sheykhi:2021fwh}. Similarly, Tsallis entropy $S_h \propto A^\beta$ \cite{Tsallis:1987eu,Lyra:1997ggy} emerges as a non-extensive generalization capable of describing gravitational systems where standard Boltzmann-Gibbs statistics fails. Remarkably, when $\beta = 1+\Delta/2$, Tsallis entropy reduces to the Barrow form, revealing a deep connection between these approaches \cite{Barrow:2020tzx,Sheykhi:2021fwh}. These entropy modifications have been extensively studied across diverse contexts including holographic dark energy \cite{Tavayef:2018xwx,Motaghi:2024rag}, inflationary cosmology \cite{Mohammadi:2021wde,Ghosh:2024jvs}, and solutions to the Hubble tension \cite{Yarahmadi:2024oqv}. 

The R\'enyi entropy \cite{renyi1961measures} is another one-parameter generalization of the Boltzmann-Gibbs entropy, originally developed in information theory. This entropy is applied to black hole thermodynamics and cosmological context \cite{Komatsu:2016vof,Moradpour:2017ycq,Naeem:2022jdq}. The modifications of the R\'enyi can influence the expansion rate of the universe, $H(t)$, particularly during critical epochs such as Big Bang Nucleosynthesis (BBN).

Kaniadakis entropy is another example of non-extensive entropy which the basic idea introduced in an effort to relativistic statistical mechanics \cite{Kaniadakis:2002zz,Kaniadakis:2005zk}. Hence the Kaniadakis entropy was emerged as \cite{Drepanou_2022,Moradpour:2020dfm,Lymperis:2021qty}. Different aspect of the Kaniadakis in to the cosmology framework is extensively investigated \cite{Lymperis:2021qty,Sheykhi:2023aqa,KordZangeneh:2023syq,P:2022amn}. Also observational constraint o Kaniadakis cosmology is discussed in \cite{zarandi2025investigating}.

Beside, people in the literature consider to other form of entropy relations in this regard. Loop quantum gravity entropy \cite{Rovelli:1996dv,Mann:1996ze,Kaul:2000kf,Das:2001ic,Banerjee:2008cf} and power law corrected form of entropy \cite{Das:2007mj,Radicella:2010ss} are two well explored examples of this field.

In this work, we investigate inflationary cosmology within the framework of extended entropies, constraining the free parameters of selected models and comparing them with the Bekenstein-Hawking (BH) case. We incorporate the method introduced by \cite{KHODAMMOHAMMADI2023138066}, which is based on defining a modified energy-momentum tensor characterized by a function $f(\rho)$ that quantifies deviations from the BH entropy (where $f=1$). The standard inflationary observables---the scalar spectral index ($n_s$) and the tensor-to-scalar ratio ($r$)---serve as primary constraints, with model parameters determined through Hamiltonian Monte Carlo (HMC) sampling. Our analysis further treats the uncertainty in the upper bound of $r$ ($\sigma_r$) as a free parameter.

The paper is structured as follows: section~\ref{sec:model} reviews the underlying theoretical framework. Section~\ref{sec:method} introduces the statistical analysis techniques and computational approach. Subsequent sections present the parameter constraints for each model and examine their distinctive properties. Section~\ref{sec:discussion} investigates the features of the inflationary potentials and its footprint in the reheating era. Also, we provide a comparative analysis of the models using statistical criteria ($\chi^2$, Gaussian tension). Finally, we summarize our conclusions.    

\section{Inflationary Model}
\label{sec:model}

In this section, we present the theoretical foundation of our analysis. The action governing the inflaton field with minimal coupling to gravity is given by:

\begin{equation}\label{eq:action}
\mathcal{S} = \int d^{4}x \sqrt{-g} \left[ \frac{1}{2}R - \Lambda + \mathcal{L}_{\phi} \right],
\end{equation}

where $\mathcal{L}_{\phi}$ is the Lagrangian density of the scalar inflaton field $\phi$. Herein and in the rest of the paper we work with the Planck units (unless explicitly noted). We adopt a spatially flat Friedmann--Robertson--Walker--Lema\^{i}tre (FRWL) metric:

\begin{equation}
ds^2 = -dt^2 + a(t)^2 h_{ij} dx^i dx^j,
\end{equation}

where $h_{ij}$ is the Euclidean metric, and assume $\Lambda = 0$. Following the work of \cite{KHODAMMOHAMMADI2023138066,khodam2024non}, a conformal transformation of the energy-momentum tensor (EMT) of the form $T_{\mu\nu} \rightarrow f(\rho)T_{\mu\nu}$ has been proposed to describe the behavior of the cosmic fluid in the context of extended entropies. This is analogous to a K-essence model with $\mathcal{L} = K(\phi)P(X)$, where the choices of $K$ and $P$ depend on $f(\rho)$ and the potential $V(\phi)$. Consequently, one may define:

\begin{equation}
\frac{1}{\sqrt{-g}} \frac{\delta \mathcal{L}_{\phi}}{\delta g^{\mu\nu}} \equiv T_{\mu\nu} =
\begin{pmatrix}
-f(\rho_{\phi})\rho_{\phi} & 0 & 0 & 0 \\
0 & f(\rho_{\phi})p_{\phi} & 0 & 0 \\
0 & 0 & f(\rho_{\phi})p_{\phi} & 0 \\
0 & 0 & 0 & f(\rho_{\phi})p_{\phi}
\end{pmatrix},
\end{equation}

for a perfect fluid evolving in a cosmic background. Note that $\rho_{\phi} = \dot{\phi}^2/2 + V(\phi)$ and $p_{\phi} = \dot{\phi}^2/2 - V(\phi)$. It is evident that the equation of state $w_{\phi} = p_{\phi}/\rho_{\phi}$ remains invariant under the aforementioned conformal transformation.

In our analysis, the function $f(\rho)$ is determined by the generalized entropy $S_G$ via the relation \cite{khodam2024non}:  

\begin{equation}
\rho f(\rho) = \frac{3}{8} S_G(S_{\text{BH}}),
\end{equation}

where $S_G$ is a function of the Bekenstein--Hawking entropy $S_{\text{BH}} = A/4 = 3/(8\rho)$. Here, $A$ denotes the area of the cosmological apparent horizon with radius $r_h = 1/H$, and $H$ is the Hubble parameter.

Several forms of $f(\rho)$ can be derived from different entropy models \cite{nojiri2022modified}. In addition to the standard Bekenstein--Hawking entropy, extended entropies can be categorized into power-law \cite{das2008power,sheykhi2023modified,sheykhi2018modified,abreu2022statistical}, logarithmic \cite{abreu2022statistical,fazlollahi2023renyi,Kaul:2000kf}, and hyperbolic \cite{abreu2022statistical,zarandi2025investigating,lymperis2021modified} types. We select one representative case from each category for our analysis. The chosen entropies and their corresponding functions $f$ are listed in table~\ref{tab:entropies}.

Varying the action \eqref{eq:action} with respect to the metric yields the modified Friedmann equations:

\begin{align}
H^2 &= \frac{8\pi}{3} \rho_{\phi} f(\rho_{\phi}), \label{eq:Friedmann1} \\
\dot{H} &= -4\pi (\rho_{\phi} + p_{\phi}) f(\rho_{\phi}). \label{eq:Friedmann2}
\end{align}

Moreover, from $\nabla_{\mu} T^{\mu\nu} = 0$, we obtain the continuity equation:

\begin{equation}
\dot{\rho}_{\phi} + 3H \rho_{\phi} (1 + w_{\phi}^{\text{eff}}) = 0, \label{eq:continuity}
\end{equation}

which incorporates an effective equation of state:

\begin{equation}
1 + w_{\phi}^{\text{eff}} = \frac{1 + w_{\phi}}{\Gamma(\rho_{\phi})}, \quad \text{where} \quad \Gamma(\rho) = 1 + \rho \frac{f'}{f}, \label{eq:Gamma}
\end{equation}

and the prime denotes differentiation with respect to $\rho$.

An alternative approach is provided by the Hamilton--Jacobi formalism \cite{lidsey1997reconstructing,tzirakis2009non,skenderis2006hamilton}, where the Hubble parameter during inflation is parametrized as an implicit function of the scalar field \cite{byrnes2009non,tzirakis2009non,skenderis2006hamilton}. This formalism allows the field potential to be directly derived from eq.~\eqref{eq:Friedmann1}. The dynamical equation for the field follows from eq.~\eqref{eq:Friedmann2}:

\begin{equation}
\dot{\phi} = -\frac{1}{4\pi} \frac{H_{,\phi}}{f}, \label{eq:field_dynamics}
\end{equation}

and the potential as a function of the field is given by:

\begin{equation}
V(\phi) = \frac{1}{8\pi^2 f^2} \left( 3\pi f H^2(\phi) + \frac{1}{4} \left( \partial_{\phi} H(\phi) \right)^2 \right). \label{eq:potential}
\end{equation}

Here, we have used eqs.~\eqref{eq:Friedmann1} and \eqref{eq:field_dynamics}. The Hubble parameter during inflation has been parametrized in various ways in the literature. Among these, the power-law \cite{tzirakis2009non,coone2015hubble,enqvist2020structure} and linear \cite{byrnes2009non,asadi2019reheating} parametrizations are of particular interest for our analysis. To accommodate greater generality, we introduce a nonlinear parametrization of the Hubble parameter:

\begin{equation}
H(\phi) = H_{\text{inf}} (h_0 + h_1 \phi^n), \label{eq:H_param}
\end{equation}

where $H_{\mathrm{inf}} \sim 10^{-5}$ (in Planck units) represents a characteristic Hubble scale during inflation, while $h_0$, $h_1$, and $n$ are real dimensionless constants. A fundamental requirement of inflation is that the Hubble parameter remains approximately constant throughout the inflationary expansion. For our parametrization in eq.~\eqref{eq:H_param}, we can demonstrate that $\partial_\phi H$ remains sufficiently small (given $h_1, n \sim \mathcal{O}(10^{-1})$ and $\phi \sim \mathcal{O}(1-10)$ (for large field inflation) \cite{kallosh2025present,vazquez2018inflationary}), thereby ensuring that the essential slow-roll conditions are satisfied. 

Within this modified framework, the slow-roll parameters are defined as:

\begin{align}
\epsilon_{\text{mod}} &= \frac{1}{16\pi} \left( \frac{V_{,\phi}}{V} \right)^2 \frac{1}{f \Gamma}, \label{eq:epsilon_mod} \\
\eta_{\text{mod}} &= \frac{1}{8\pi} \frac{V_{,\phi\phi}}{V} \frac{1}{f}. \label{eq:eta_mod}
\end{align}

The modified power spectrum of scalar perturbations, $\mathcal{P}_{\mathcal{R}} \propto H^2 / \epsilon_{\text{mod}}$, evaluated at the horizon crossing scale $k = aH$ (corresponding to the field value $\phi_0$), yields the spectral index \cite{khodam2024non}:

\begin{equation}
n_s = 1 + \frac{d \ln \mathcal{P}_{\mathcal{R}}}{d \ln k} \approx 1 - 6 \epsilon_{\text{mod}} + 2 \eta_{\text{mod}}. \label{eq:ns}
\end{equation}

Similarly, the power spectrum of tensor perturbations, $\mathcal{P}_t \propto H^2$, leads to the tensor-to-scalar ratio:

\begin{equation}
r = \frac{\mathcal{P}_t}{\mathcal{P}_{\mathcal{R}}} = 16 \epsilon_{\text{mod}}. \label{eq:r}
\end{equation}

By evaluating $V$, $V'$, and $V''$ at $\phi = \phi_0$ in eqs.~\eqref{eq:epsilon_mod} and \eqref{eq:eta_mod}, the parameters $n_s$ and $r$ can be expressed in terms of the model parameters.

The slow-roll conditions, $\epsilon \ll 1$ and $|\eta| \ll 1$, ensure a prolonged inflationary expansion, resolving the horizon problem (see appendix~\ref{app:slow-roll} for validation of slow-roll conditions). Inflation ends when these conditions are violated, i.e., when $\epsilon_{\text{mod}} \sim 1$ \cite{vazquez2018inflationary}. The scalar field value $\phi_e$ at the end of inflation is determined by the requirement that the universe expands by $N_e$ e-folds:

\begin{equation}
N_e = \int_{t_0}^{t_e} H(t) dt = 4\pi \int_{\phi_e}^{\phi_0} f \frac{H}{H_{,\phi}} d\phi. \label{eq:e-folds}
\end{equation}
   
\begin{table}[tbp]
\centering
\caption{Entropy models and their corresponding functions}
\label{tab:entropies}
\begin{tabular}{lcc}
\toprule
Model & Entropy Relation & $f(\rho)$ Function \\
\midrule
Bekenstein-Hawking entropy (BH) & $S_{BH}$ & $1$ \\
Tsallis entropy (T) & $S_0 \left( \dfrac{S_{BH}}{S_0} \right)^{\delta}$ & $\left( \dfrac{\rho}{\rho_0} \right)^{\delta - 1}$ \\
R\'enyi entropy (R) & $\dfrac{1}{\alpha}\ln(\alpha S_{BH} + 1)$ & $\dfrac{3\alpha}{8\rho} \left[\ln\left(1 + \dfrac{3\alpha}{8\rho}\right)\right]^{-1}$ \\
Kaniadakis entropy (K) & $\dfrac{1}{K}\sinh(K S_{BH})$ & $\dfrac{3K}{8\rho} \left[\sinh\left(\dfrac{3K}{8\rho}\right)\right]^{-1}$ \\
\bottomrule
\end{tabular}
\end{table}

\section{Statistical Analysis}
\label{sec:method}

We examine inflationary models based on extended entropies by comparing their predictions with observational data. In this way, we employ the HMC method using the \texttt{PyMC} package, which provides efficient sampling through the No-U-Turn Sampler (NUTS) algorithm \cite{salvatier2016probabilistic,abril2023pymc}. We perform $2.4 \times 10^4$ draws for each model and verify convergence using the diagnostics $\hat{R} \leq 1.01$ and $\text{ESS}_{\text{bulk}}, \text{ESS}_{\text{tail}} \gtrsim 1000$, ensuring the robustness of our results \cite{vehtari2021rank}.

The primary observational constraints on inflationary parameters are the scalar spectral index $n_{s}$ and the tensor-to-scalar ratio $r$. Current measurements report $n_{s} = 0.974 \pm 0.003$ \cite{louis2025atacama,calabrese2025atacama} and $r < 0.038$ (in 95\% confidence level) \cite{calabrese2025atacama}. The theoretical expressions for these parameters are given in terms of the modified slow-roll parameters $\epsilon_{\text{mod}}$ and $\eta_{\text{mod}}$.

In the Bayesian framework, the posterior probability distribution is given by $P(\Theta | D) \propto \mathcal{L}(D|\Theta)P(\Theta)$, where $\mathcal{L}$ is the likelihood function and $P(\Theta)$ represents the prior probability distribution. Here, $\Theta$ and $D$ denote the sets of free parameters and observational data, respectively. We specify $P(\Theta)$ for each model in subsequent sections. The definition of the likelihood functions is the hallmark of our analysis.

For $n_{s}$, we adopt a Gaussian likelihood:
\begin{equation}\label{eq:likens}
\mathcal{L}_{n_{s}} \propto \exp\left[-\frac{(n_{s}(\Theta)-n_{s}^{\star})^{2}}{2\sigma_{n_{s}}^{2}}\right],
\end{equation}
where $n_{s}^{\star} = 0.974$ and $\sigma_{n_{s}} = 0.003$. 

The situation for $r$ is different, as we only have an upper bound $r^{\star} = 0.038$. We consider the piece-wise likelihood functions to handle this constraint:
\begin{equation}\label{eq:liker1}
\mathcal{L}_{r1} \propto
\begin{cases} 
1 & 0 < r < r^{\star} \\
\exp\left[-\frac{(r(\Theta)-r^{\star})^{2}}{2(0.001)^{2}}\right] & r \geq r^{\star},
\end{cases}
\end{equation}
and
\begin{equation}\label{eq:liker2}
\mathcal{L}_{r2} \propto
\begin{cases} 
1 & 0 < r < r^{\star} \\
\exp\left[-\frac{(r(\Theta)-r^{\star})^{2}}{2\sigma_{r}^{2}}\right] & r \geq r^{\star}.
\end{cases}
\end{equation}

From the normalized likelihood
\begin{equation}
P(r<r^{\star})=\frac{r^{\star}}{r^{\star}+\sigma_{r}\sqrt{\tfrac{\pi}{2}}},
\end{equation}
imposing $P(r<0.038)=0.95$ yields $\sigma_{r}=r^{\star}(1-P)/P\sqrt{\pi/2}\simeq 1.6\times 10^{-3}$.
Thus, at the 95\% confidence level, the width parameter is validated to be
\(\sigma_{r}\sim \mathcal{O}(10^{-3})\). In $\mathcal{L}_{r1}$, we fix the uncertainty in $r$ to $10^{-3}$, while in $\mathcal{L}_{r2}$, $\sigma_{r}$ is treated as a free parameter to accommodate potentially larger values of $r$. We refer to models using $\mathcal{L}_{r1}$ and $\mathcal{L}_{r2}$ as Model 1 and Model 2, respectively. A similar likelihood is proposed for ages of the oldest astrophysical objects \cite{vagnozzi2022implications,costa2023bias}.

\section{Bekenstein-Hawking Entropy}

Using eq.~\eqref{eq:Friedmann1} and extracting $f_{\text{BH}}$ from table~\ref{tab:entropies}, one obtains the energy density of the inflationary field as:
\begin{equation}
\rho_{\phi}^{(\text{BH})} = \frac{3H^{2}(\phi)}{8\pi}.
\label{eq:rho_bh}
\end{equation}

Over time, as the inflationary field decreases, the expansion rate decreases slightly as described by eq.~\eqref{eq:field_dynamics}. Consequently, the energy density of the scalar field decays until the slow-roll conditions are ultimately broken at the end of inflation. This behavior aligns with the standard inflationary scenario. To validate this picture, we set the free parameter values accordingly.
In particular, we adopt uniform prior for all model parameters.  
The parameters $h_0$, $h_1$, and $n$ are restricted to the interval $(0, 1)$, while $\phi_0$ is defined within the range $(1, 10)$.  
These choices ensure that $H > 0$, $H_{\phi} > 0$, and $\dot{\phi} < 0$ throughout the evolution.
Although values such as $n < 0$ are mathematically allowed, they lead to physically disfavored scenarios.  
We therefore exclude such cases from our analysis.

\subsection{Observational Results of BH Models}\label{secBH}

The results for BH-1 model and its extension (BH-2) are summarized in table~\ref{tab:cosmology_fit_results_BH}. The BH-2 model introduces an additional parameter $\sigma_r$ compared to BH-1, accounting for uncertainty in measurement of $r$. Both BH-1 and BH-2 yield consistent parameter estimates, with $\sigma_r \simeq 0.006$ in BH-2 case. The common parameters show slight difference between two models, particularly in $\phi_0$ ($\simeq 1.20$ for BH-1 and $\simeq 1.27$ for BH-2) and $n$ ($\simeq 0.31$ for BH-1 and $\simeq 0.35$ for BH-2).

The uncertainty in the tensor-to-scalar ratio, $\sigma_r$, determines the upper bound on $r$.  
Since $\sigma_r$ is larger in the BH-2 case, the allowed upper limit on $r$ also becomes slightly higher.  
Moreover, due to degeneracies between the parameters $\phi_0$--$r$ and $n$--$r$, slightly larger values of $\phi_0$ and $n$ are permitted in the BH-2 scenario. Regarding the value of $n$, it is also worth noting that, from eq.~\eqref{eq:potential}, a value of $n \sim 0.3$ corresponds to a potential $V \sim \phi^{0.6}$, which is consistent with recent ACT results \cite{AtacamaCosmologyTelescope:2025nti}.

The derived cosmological parameters for both BH models are presented in table~\ref{tab:cosmology_fit_results_BH}. For a direct comparison, we evaluate the well-known Starobinsky model predictions \cite{mukhanov1981quantum} for a matching effective e-fold number of $N_e \simeq 49$ (derived for both models), yielding $\tilde{n}_s \simeq 1 - 2/N_e \simeq 0.959$ and $\tilde{r} \simeq 12/N_e^2 \simeq 0.005$. Our models consistently produce higher values ($n_s \simeq 0.976-0.977, r \simeq 0.033-0.037$). While the elevated $n_s$ suggests a modification to the inflationary potential, the more pronounced enhancement in $r$, particularly for BH-2 model, points toward a parameterization that inherently allows for a stronger primordial gravitational wave signal.

\begin{figure}[tbp]
\centering
\includegraphics[width=0.96\linewidth]{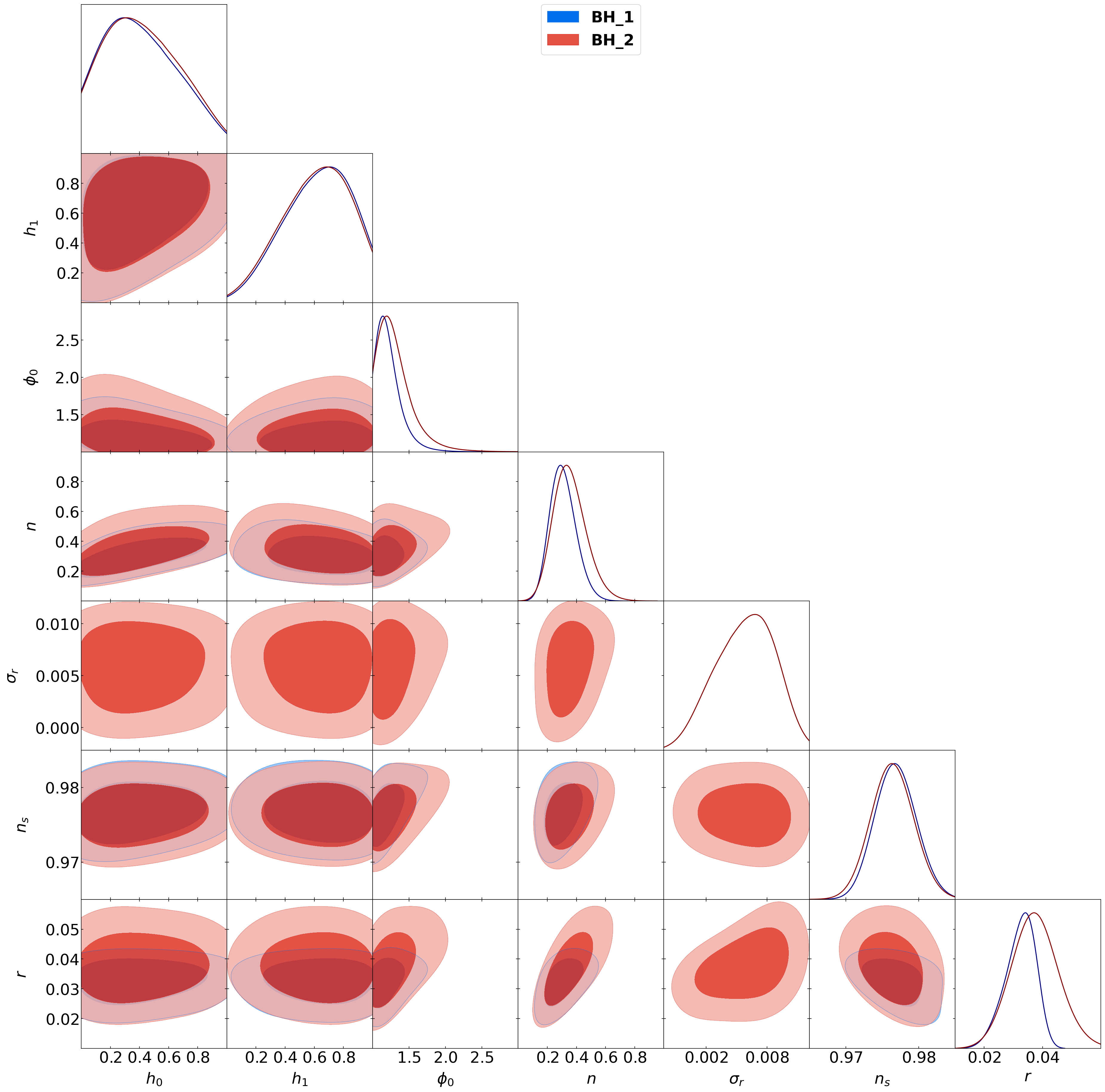}
\caption{BH-1 and BH-2 models parameter distributions and correlations}
\label{fig:BH_results}
\end{figure}

The correlation structures in figure~\ref{fig:BH_results} show that $\sigma_r$ is largely independent of other parameters, indicating it captures distinct physical effects. 

\begin{table}[tbp]
\centering
\caption{Parameters for BH models}
\label{tab:cosmology_fit_results_BH}
\begin{tabular}{lcc}
\toprule
Parameter & BH-1 & BH-2 \\
\midrule
$h_0$ & $0.393^{+0.132}_{-0.392}$ & $0.407^{+0.132}_{-0.407}$ \\
$h_1$ & $0.633^{+0.365}_{-0.117}$ & $0.629^{+0.368}_{-0.124}$ \\
$\phi_0$ & $1.198^{+0.037}_{-0.198}$ & $1.269^{+0.048}_{-0.269}$ \\
$n$ & $0.309^{+0.067}_{-0.102}$ & $0.352^{+0.073}_{-0.134}$ \\
$\sigma_r$ & - & $0.006^{+0.004}_{-0.002}$ \\
\midrule
$n_s$ & $0.977^{+0.002}_{-0.003}$ & $0.976^{+0.004}_{-0.004}$ \\
$r$ & $0.033^{+0.006}_{-0.008}$ & $0.037^{+0.009}_{-0.012}$ \\
\bottomrule
\end{tabular}
\end{table}

\section{Tsallis Entropy}

Using eq.~\eqref{eq:Friedmann1} and the expression for $f_{T}$ from table~\ref{tab:entropies}, the energy density of the Tsallis scalar field is given by:
\begin{equation}
\rho_{\phi}^{(T)} = \frac{3H_{0}^{2}}{8\pi}\left( \frac{H}{H_{0}} \right)^{\frac{2}{\delta}}.
\label{eq:rho_tsallis}
\end{equation}
Here, we have used the ansatz $S_{T}(0) = S_{\text{BH}}(0)$. In the limit $\delta \rightarrow 1$, the standard Bekenstein-Hawking energy density $\rho_{\phi}^{(\text{BH})}$ is recovered.

In general, the relation $\rho_{\phi}^{(T)} \propto \left( \rho_{\phi}^{(\text{BH})} \right)^{\frac{1}{\delta}}$ is obtained, where the Tsallis parameter $\delta$ acts as an intermediate constant that adjusts the strength of the inflationary field's energy density. For consistency with the slow-roll inflation scenario, we require $\delta > 0$. Conditions $\delta < 1$ and $\delta > 1$ lead to lower and higher energy densities for Tsallis inflation, respectively (note that $\rho_{\phi} < 1$ in our units).

\subsection{Observational Results of T Models}

The results for the T-1 model and its extension (T-2) are summarized in table~\ref{tab:cosmology_fit_results_T}. The T-2 model introduces an additional parameter, $\sigma_r$, to account for measurement uncertainty in the tensor-to-scalar ratio $r$. Overall, both models yield consistent parameter estimates, with the derived $\sigma_r \simeq 0.006$ for T-2. The shared parameters show slight variations between the models, most notably in $\phi_0$ ($\simeq 1.19$ for T-1 vs. $\simeq 1.34$ for T-2) and $\delta$ ($\simeq 1.26$ for T-1 vs. $\simeq 1.16$ for T-2), where the prior for $\delta$ was set to the interval (0,2) \cite{teimoori2024inflation}.

The derived cosmological parameters for both T models are presented in table~\ref{tab:cosmology_fit_results_T}. For a direct comparison, we evaluate the standard Starobinsky predictions using the effective e-fold numbers derived from each model, $N_e \simeq 57$ for T-1 and $N_e \simeq 46$ for T-2. This yields benchmark values of $(\tilde{n}_s, \tilde{r}) \simeq (0.965, 0.004)$ for T-1 and $(0.956, 0.006)$ for T-2. In contrast, our T models consistently produce higher values, with $n_s \simeq 0.976$--$0.977$ and $r \simeq 0.032$--$0.037$. While the elevated $n_s$ suggests a modification to the inflationary potential, the more pronounced enhancement in $r$--particularly for T-2--indicates a parameterization that inherently favors stronger production of primordial gravitational waves.

\begin{table}[tbp]
\centering
\caption{Parameters for T models}
\label{tab:cosmology_fit_results_T}
\begin{tabular}{lcc}
\toprule
Parameter & T-1 & T-2 \\
\midrule
$h_0$ & $0.325^{+0.087}_{-0.325}$ & $0.354^{+0.103}_{-0.354}$ \\
$h_1$ & $0.634^{+0.366}_{-0.117}$ & $0.620^{+0.380}_{-0.123}$ \\
$\phi_0$ & $1.194^{+0.031}_{-0.194}$ & $1.344^{+0.034}_{-0.344}$ \\
$n$ & $0.383^{+0.154}_{-0.197}$ & $0.398^{+0.173}_{-0.212}$ \\
$\delta$ & $1.258^{+0.742}_{-0.239}$ & $1.164^{+0.707}_{-0.382}$ \\
$\sigma_r$ & - & $0.006^{+0.004}_{-0.002}$ \\
\midrule
$n_s$ & $0.977^{+0.003}_{-0.004}$ & $0.976^{+0.004}_{-0.004}$ \\
$r$ & $0.032^{+0.007}_{-0.008}$ & $0.037^{+0.009}_{-0.013}$ \\
\bottomrule
\end{tabular}
\end{table}

\begin{figure}[tbp]
\centering
\includegraphics[width=0.96\linewidth]{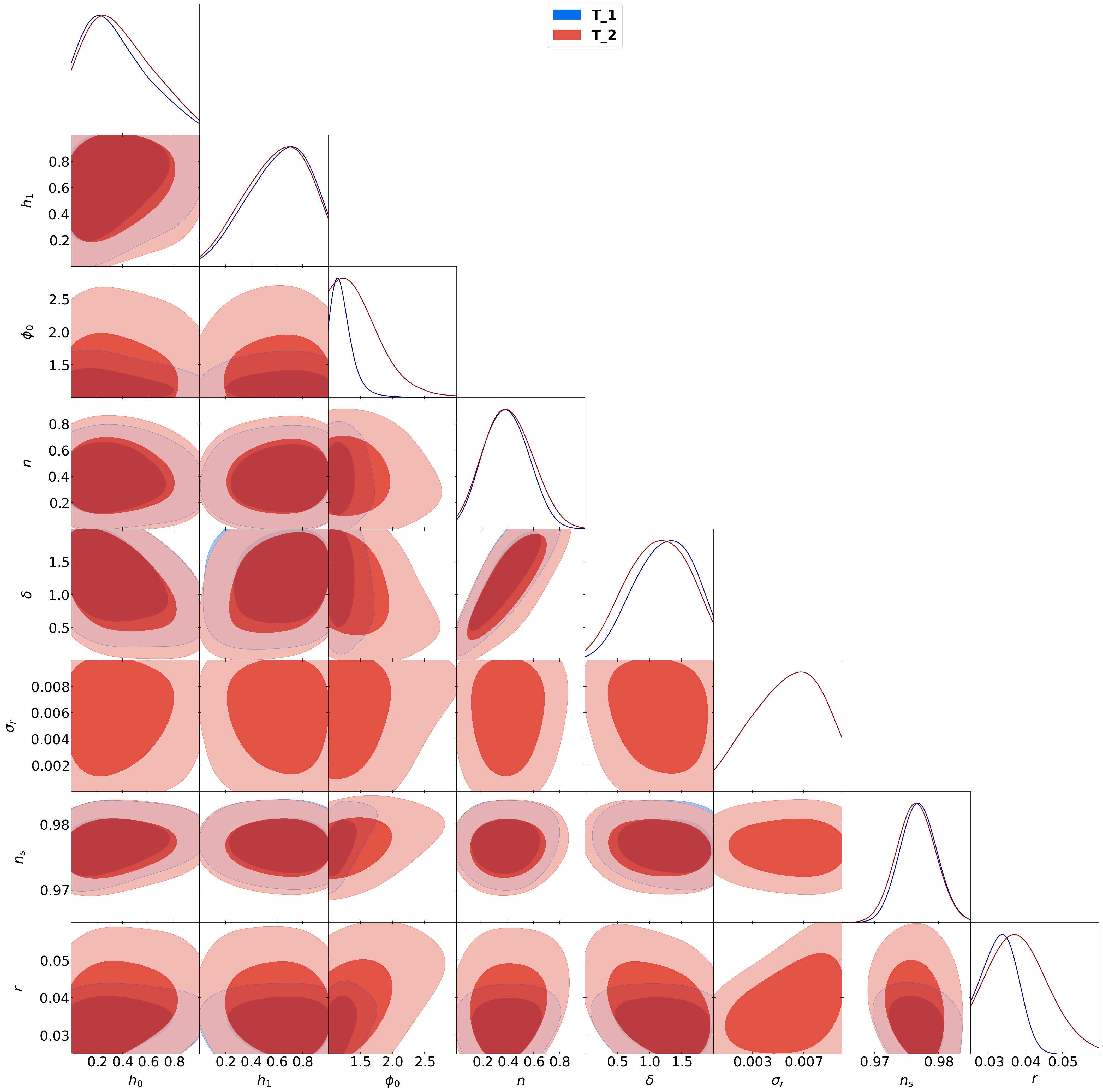}
\caption{T-2 model parameter distributions and correlations}
\label{fig:T_results}
\end{figure}

\section{R\'enyi Entropy}

Substituting $f_{R}$ from table~\ref{tab:entropies} into eq.~\eqref{eq:Friedmann1}, we obtain the energy density for the R\'enyi scalar field:
\begin{equation}
\rho_{\phi}^{(R)} = \frac{3}{8}\frac{\alpha}{e^{\frac{\pi\alpha}{H^{2}}} - 1},
\label{eq:rho_renyi}
\end{equation}
The Bekenstein-Hawking entropy is recovered in the limit $\alpha \rightarrow 0$. In this limit, $\rho_{\phi}^{(R)} \propto \rho_{\phi}^{(\text{BH})}$, implying $\pi\alpha/H^{2} \ll 1$. However, in our analysis, various values of $\alpha$ may lead to a perturbed energy density for R\'enyi models relative to the BH case, expressed as $\rho_{\phi}^{(R)} \propto \rho_{\phi}^{(\text{BH})} + \alpha\delta\rho_{\phi}^{(\text{BH})} + \ldots$.

\subsection{Observational Results of R Models}

The results for the R-1 model and its extension (R-2) are summarized in table~\ref{tab:cosmology_fit_results_R}. The R-2 model incorporates an additional parameter, $\sigma_r$, to account for measurement uncertainty in the tensor-to-scalar ratio $r$. Both models yield consistent parameter estimates, with $\sigma_r \simeq 0.006$ derived for R-2. The common parameters show minor variations between the models, most notably in $\phi_0$ ($\simeq 1.20$ for R-1 vs. $\simeq 1.31$ for R-2) and $n$ ($\simeq 0.31$ for R-1 vs. $\simeq 0.35$ for R-2). The prior for $-\log\alpha$ was set to the interval $(8,18)$ \cite{khodam2024non}.

The derived cosmological parameters for both R models are presented in table~\ref{tab:cosmology_fit_results_R}. For direct comparison, we evaluate the standard Starobinsky predictions using the effective e-fold numbers derived from each model, $N_e \simeq 50$ for R-1 and $N_e \simeq 53$ for R-2. This yields benchmark values of $(\tilde{n}_s, \tilde{r}) \simeq (0.960, 0.005)$ for R-1 and $(0.962, 0.004)$ for R-2. In contrast, our R models consistently produce higher values, with $n_s \simeq 0.976$--$0.977$ and $r \simeq 0.033$--$0.038$. While the elevated $n_s$ suggests a modification to the inflationary potential, the more pronounced enhancement in $r$--particularly for R-2--indicates a parameterization that inherently favors stronger primordial gravitational wave production.

\begin{table}[tbp]
\centering
\caption{Parameters for R models}
\label{tab:cosmology_fit_results_R}
\begin{tabular}{lcc}
\toprule
Parameter & R-1 & R-2 \\
\midrule
$h_0$ & $0.409^{+0.138}_{-0.409}$ & $0.423^{+0.143}_{-0.423}$ \\
$h_1$ & $0.644^{+0.345}_{-0.119}$ & $0.640^{+0.359}_{-0.114}$ \\
$\phi_0$ & $1.197^{+0.036}_{-0.197}$ & $1.312^{+0.012}_{-0.312}$ \\
$n$ & $0.306^{+0.069}_{-0.099}$ & $0.351^{+0.079}_{-0.135}$ \\
$-\log\alpha$ & $14.378^{+3.596}_{-1.311}$ & $14.207^{+3.624}_{-1.519}$ \\
$\sigma_r$ & - & $0.006^{+0.004}_{-0.001}$ \\
\midrule
$n_s$ & $0.977^{+0.002}_{-0.004}$ & $0.976^{+0.004}_{-0.004}$ \\
$r$ & $0.033^{+0.006}_{-0.008}$ & $0.038^{+0.008}_{-0.013}$ \\
\bottomrule
\end{tabular}
\end{table}

\begin{figure}[tbp]
\centering
\includegraphics[width=0.96\linewidth]{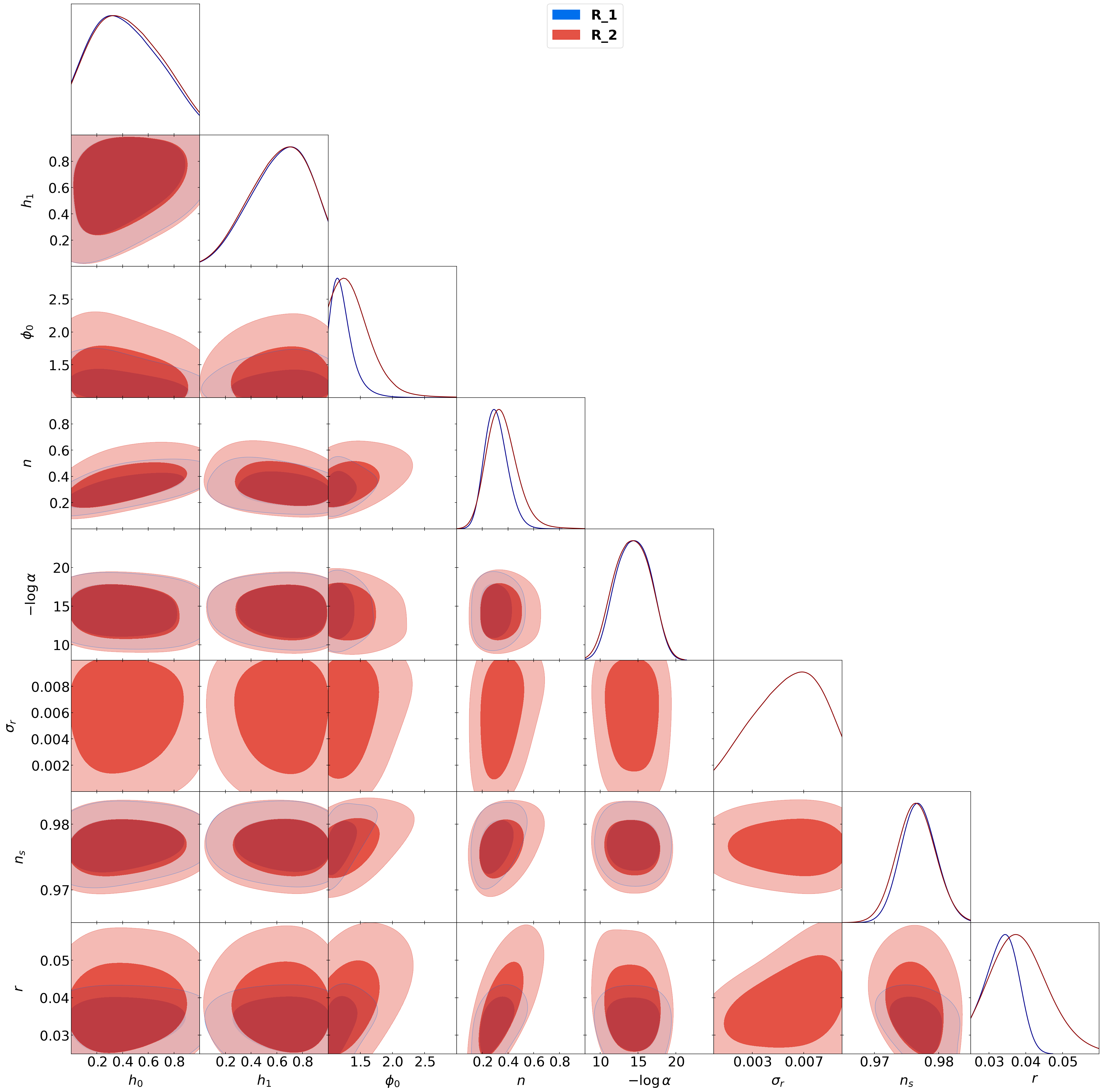}
\caption{R-2 model parameter distributions and correlations}
\label{fig:R_results}
\end{figure}

\section{Kaniadakis Entropy}

Extracting $f_{K}$ from table~\ref{tab:entropies} and substituting into eq.~\eqref{eq:Friedmann1}, we obtain the energy density for the Kaniadakis scalar field:
\begin{equation}
\rho_{\phi}^{(K)} = \frac{3}{8}\frac{K}{\sinh^{-1}\left( \frac{\pi K}{H^{2}} \right)},
\label{eq:rho_kaniadakis}
\end{equation}
The Bekenstein-Hawking limit is recovered as $K \rightarrow 0$. Analogous to the previous case, $\rho_{\phi}^{(K)} \propto \rho_{\phi}^{(\text{BH})}$ under the condition $\pi K/H^{2} \ll 1$. We consider different $K$ values to detect potential deviations of $\rho_{\phi}^{(K)}$ from the BH case.

\subsection{Observational Results of K Models}

The parameter estimates for the K-1 model and its extension, K-2 (which includes an additional parameter $\sigma_r$ to model uncertainty in $r$), are compiled in table~\ref{tab:cosmology_fit_results_K}. The models return consistent results, with $\sigma_r \simeq 0.006$ in K-2. Minor variations are observed in the shared parameters; especially, $\phi_0$ increases from $\simeq 1.19$ in K-1 to $\simeq 1.27$ in K-2, while $n$ shifts from $\simeq 0.31$ to $\simeq 0.35$. The prior for $-\log K$ was set over the interval $(12,22)$ \cite{lambiase2023slow}.

The corresponding cosmological predictions are listed in table~\ref{tab:cosmology_fit_results_K}. Both K models yield an effective e-fold number of $N_e \simeq 49$. For this $N_e$, the canonical Starobinsky model predicts $(\tilde{n}_s, \tilde{r}) \simeq (0.959, 0.005)$. Our results, with $n_s \simeq 0.976$--$0.977$ and $r \simeq 0.033$--$0.037$, systematically exceed these benchmarks. This elevation in $n_s$ implies a modified inflationary potential, whereas the more substantial increase in $r$---most evident in K-2---signals a mechanism that amplifies the primordial gravitational wave background relative to the standard scenario.

\begin{figure}[tbp]
\centering
\includegraphics[width=0.96\linewidth]{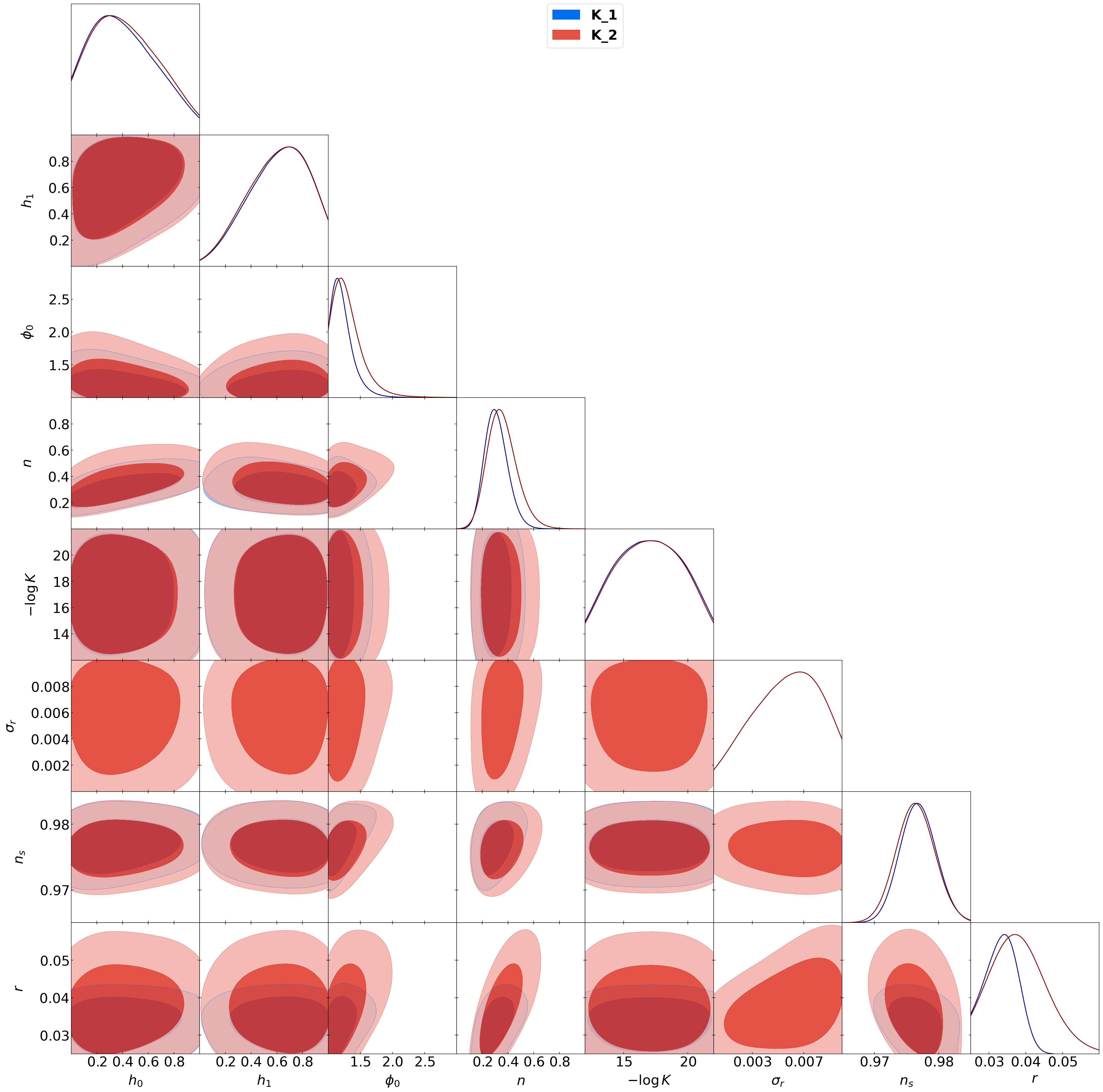}
\caption{K-2 model parameter distributions and correlations}
\label{fig:K_results}
\end{figure}

Analysis of the correlation structures presented in figure~\ref{fig:K_results} further reveals that $\sigma_r$ exhibits minimal correlation with the other parameters, affirming its role in isolating a specific source of observational uncertainty.

\begin{table}[tbp]
\centering
\caption{Parameters for K models}
\label{tab:cosmology_fit_results_K}
\begin{tabular}{lcc}
\toprule
Parameter & K-1 & K-2 \\
\midrule
$h_0$ & $0.389^{+0.124}_{-0.389}$ & $0.405^{+0.135}_{-0.405}$ \\
$h_1$ & $0.634^{+0.366}_{-0.113}$ & $0.628^{+0.372}_{-0.118}$ \\
$\phi_0$ & $1.195^{+0.035}_{-0.195}$ & $1.266^{+0.050}_{-0.266}$ \\
$n$ & $0.306^{+0.061}_{-0.106}$ & $0.352^{+0.082}_{-0.126}$ \\
$-\log K$ & $17.025^{+4.940}_{-1.845}$ & $17.017^{+3.944}_{-2.739}$ \\
$\sigma_r$ & - & $0.006^{+0.004}_{-0.002}$ \\
\midrule
$n_s$ & $0.977^{+0.002}_{-0.004}$ & $0.976^{+0.004}_{-0.004}$ \\
$r$ & $0.033^{+0.006}_{-0.009}$ & $0.037^{+0.009}_{-0.012}$ \\
\bottomrule
\end{tabular}
\end{table}

\section{Discussion and Comparison}
\label{sec:discussion}

\begin{figure}[tbp]
\centering
\includegraphics[width=0.96\linewidth]{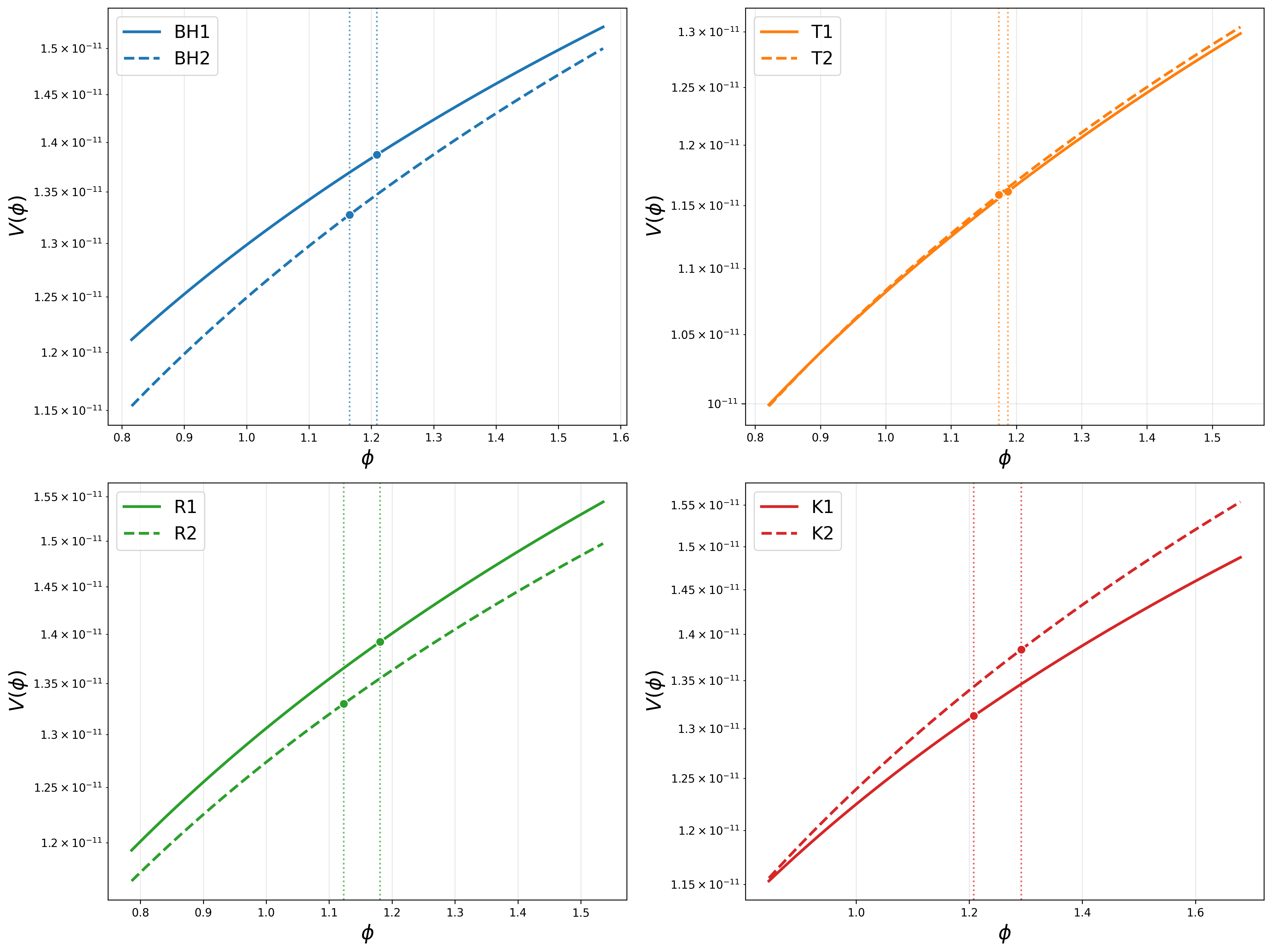}
\caption{Inflation potentials for different entropy models using best-fit parameters from table~\ref{tab:best}. The vertical dashed lines in each panel intersect the curves at their respective $\phi_0$ values.}
\label{fig:potentials}
\end{figure}

In the previous section, we reviewed the observational results on the free parameters of all four models. In the present section, the best fits of free parameters (table~\ref{tab:best}), some features of the models and corresponding physical parameters are presented. First, we consider inflation potential as a hallmark of the models. One can write down:
\begin{equation}
V^{(I)}(\phi) \simeq \rho_{\phi}^{(I)},
\label{eq:potential_approx}
\end{equation}
where $I \in \{\text{BH}, \text{T}, \text{R}, \text{K}\}$ (see appendix~\ref{app:potential} for details). The potential of models are plotted in figure~\ref{fig:potentials} considering the best fit values from table~\ref{tab:best} for free parameters. As one can see, the generalized entropy models exhibit modified potential landscapes compared to the standard Bekenstein-Hawking case, leading to altered inflationary dynamics and observable predictions. As discussed for the BH model in section~\ref{secBH}, since $n\simeq 0.3$, one obtains $V(\phi)\propto \phi^{0.6}$, which is in agreement with recent ACT observations \cite{AtacamaCosmologyTelescope:2025nti}.

\begin{figure}[tbp]
\centering
\includegraphics[width=0.96\linewidth]{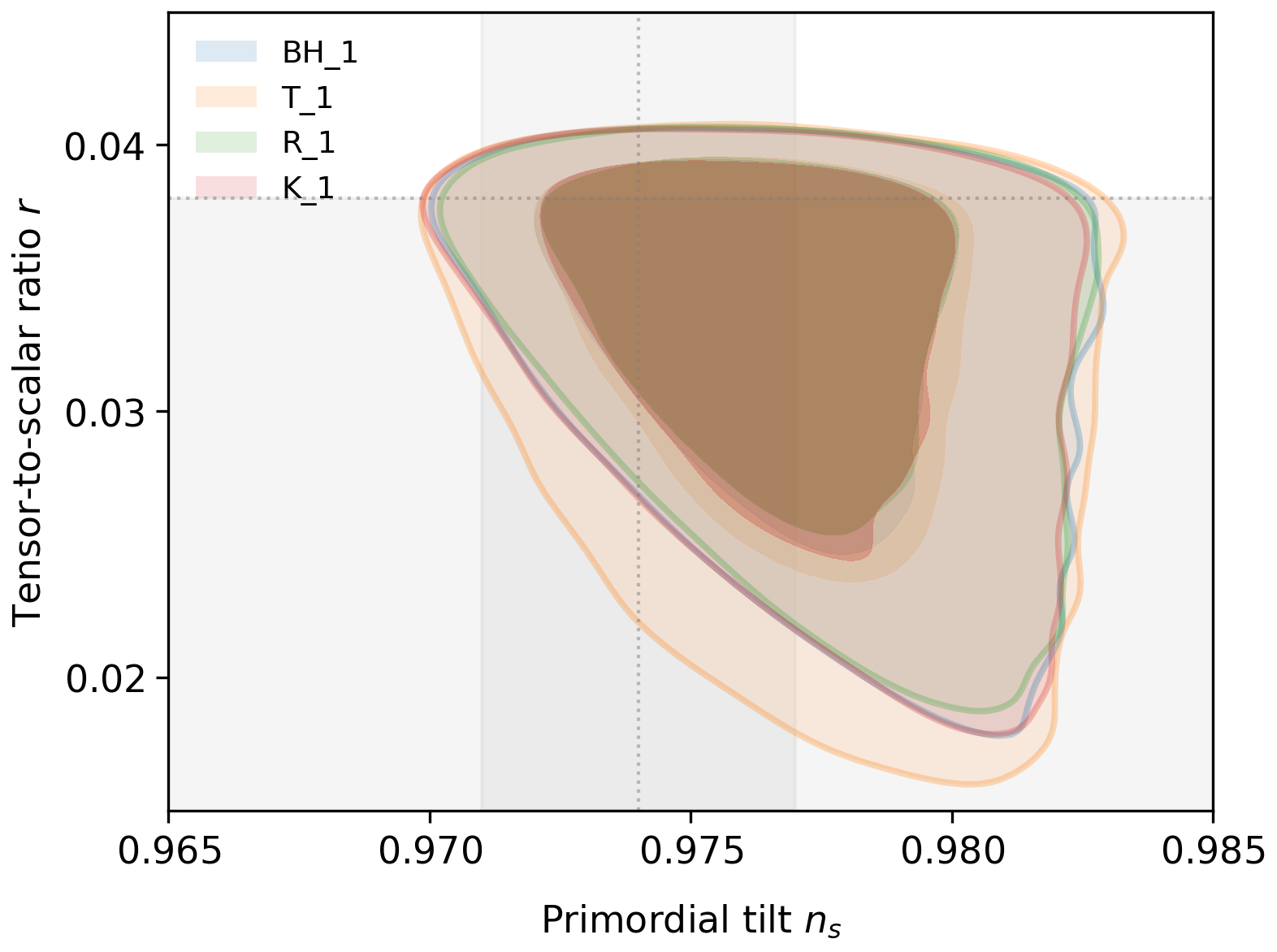}
\caption{$n_{s}-r$ diagram for models 1}
\label{fig:ns-r1}

\includegraphics[width=0.96\linewidth]{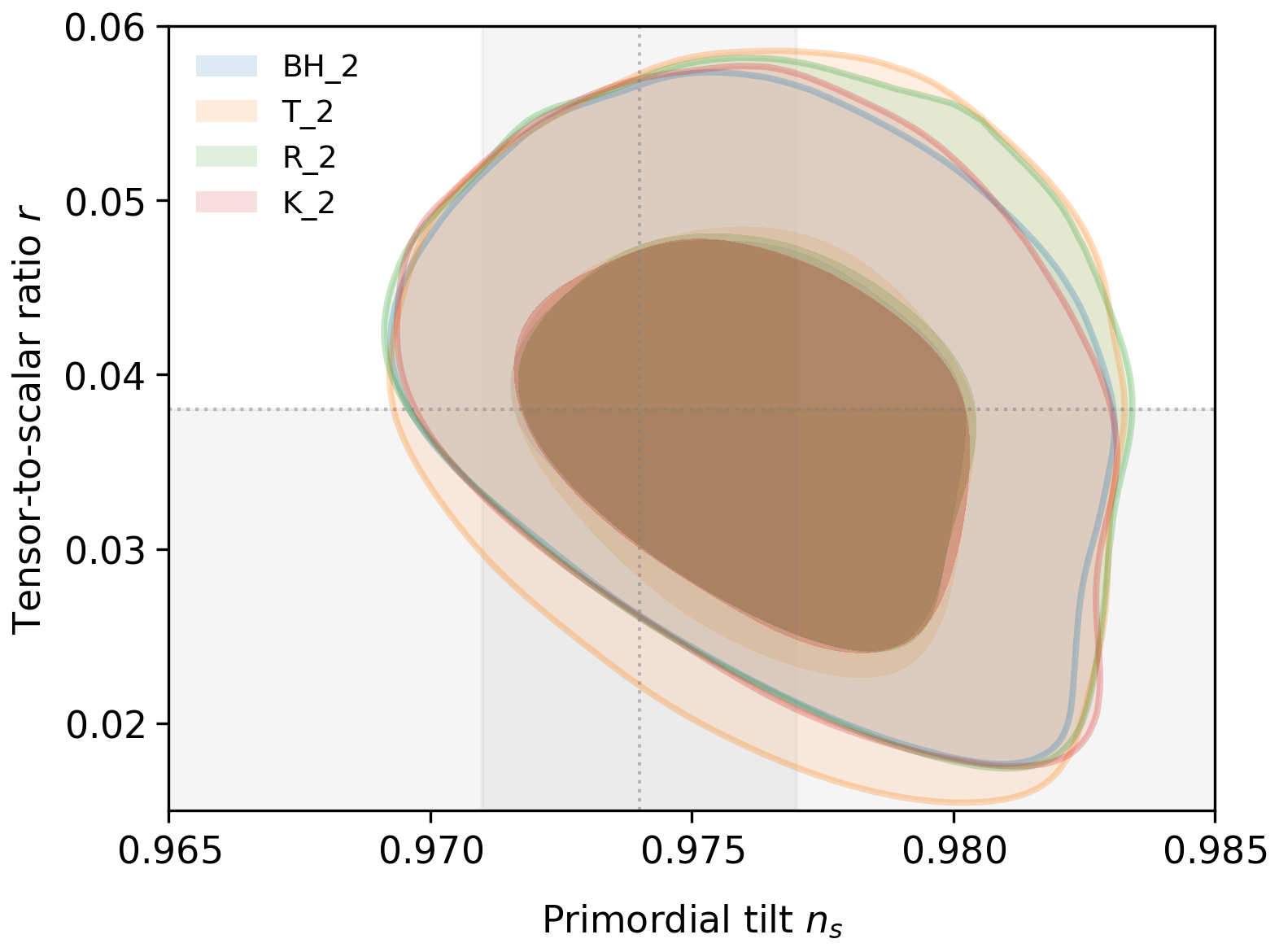}
\caption{$n_{s}-r$ diagram for models 2}
\label{fig:ns-r2}
\end{figure}

To quantitatively compare the performance of different entropy models, we employ some statistical criteria including $\chi^2$ that we used $\chi_{min}^2=-2ln(\mathcal{L}_{max})$ by definition, along with Gaussian Tension (GT) \cite{schoneberg2022h0}. We incorporate both of them for $n_{s}$, Because they are not well-defined for non-Gaussian likelihoods of $r$. In this manner, one reads: 
\begin{equation}
GT_{n_{s}} = \frac{\vert 0.974 - n_{s}(\Theta) \vert}{\sqrt{(0.003)^{2}+0.5(\sigma_{+}^{2}+\sigma_{-}^{2})}},
\label{eq:GT}
\end{equation}
In which $n_{s}(\Theta)^{+\sigma_{+}}_{-\sigma_{-}}$ is the model-derived quantity. The results of these comparisons are summarized in table~\ref{tab:model_diag}. As can be seen, the BH-2 and K-2 models exhibit minimal tension relative to the pivotal value $n_s^\star$, while the T models show a more significant deviation from it. Although the minimum $\chi^2$ value is not statistically decisive here (as we expect $\chi^2_{\mathrm{min}} \simeq 0$ for all models), one can nevertheless conclude that the BH-1 model does not provide an efficient fit, whereas the BH-2 and R-2 models perform the task perfectly. Overall, the species-2 models---which accommodate a broader allowed range for $r$---appear more promising in this context. This suggests a theoretical preference for scenarios predicting a more detectable tensor-to-scalar ratio (see also figure~\ref{fig:ns-r2}).

Figures~\ref{fig:ns-r1}--\ref{fig:ns-r2} indicate that modified entropy formulations lead to several physical consequences. Tsallis entropy produces the most significant deviations from pivotal BH case, with enhanced energy density and altered perturbation spectra. R\'enyi entropy shows intermediate behavior, with small but non-zero $\alpha$ values preferred by data. Kaniadakis entropy yields the most conservative modifications, remaining close to Bekenstein-Hawking predictions.

All non-extensive entropy models produce enhanced values of both $n_s$ and $r$ compared to pure Starobinsky $R^2$ model predictions, suggesting these generalized frameworks allow for broader inflationary scenarios.

\begin{table}[tbp]
\centering
\caption{Best-fit parameters for all models}
\label{tab:best}
\scriptsize
\begin{tabular}{lccccccccccc}
\toprule
Model & $h_0$ & $h_1$ & $\phi_0$ & $n$ & $\delta$ & $-\log\alpha$ & $-\log K$ & $\sigma_r$ & $n_s$ & $r$ & $N_e$ \\
\midrule
BH1 & 0.409 & 0.634 & 1.209 & 0.284 & - & - & - & - & 0.979 & 0.027 & 55.167 \\
BH2 & 0.365 & 0.658 & 1.165 & 0.307 & - & - & - & 0.0099 & 0.974 & 0.038 & 44.919 \\
\midrule
T1 & 0.318 & 0.622 & 1.187 & 0.389 & 1.259 & - & - & - & 0.977 & 0.032 & 36.008 \\
T2 & 0.303 & 0.644 & 1.173 & 0.345 & 1.121 & - & - & 0.0099 & 0.974 & 0.038 & 38.289 \\
\midrule
R1 & 0.412 & 0.634 & 1.181 & 0.314 & - & 14.372 & - & - & 0.976 & 0.035 & 48.066 \\
R2 & 0.356 & 0.677 & 1.123 & 0.282 & - & 14.207 & - & 0.0024 & 0.974 & 0.036 & 44.556 \\
\midrule
K1 & 0.393 & 0.620 & 1.208 & 0.297 & - & - & 17.020 & - & 0.978 & 0.030 & 52.406 \\
K2 & 0.389 & 0.630 & 1.292 & 0.341 & - & - & 17.014 & 0.0001 & 0.977 & 0.037 & 51.476 \\
\bottomrule
\end{tabular}
\end{table}

A well-studied class of inflationary scenarios is given by polynomial potentials of the form $V(\phi) \propto \phi^{\,p}$, which includes chaotic inflation for $p=2$. In the slow-roll approximation, such models predict a scalar spectral index $n_s$ and a tensor-to-scalar ratio $r$ that depend on the exponent $p$ and the number of e-folds $N_e$ as $n_s = 1 - (2+p)/(2N_e)$ and $r = 4p/N_e$ \cite{vazquez2018inflationary}.  
Using the best-fit values of $n_s$, $r$, and $N_e$ obtained from table~\ref{tab:best}, we compute the effective exponent $p$ inferred from each relation. While the $r$--$N_e$ relation yields $p \simeq 0.4$ for all of models (but T1, which gives $p \simeq 0.3$), the $n_s$--$N_e$ relation returns $p \approx 0.4$ only for the BH-2 and K-2 scenarios. For the remaining models, in particular the T models, the two estimates of $p$ differ significantly, indicating a departure from the simple power-law form.  
Consequently, the BH-2 and K-2 best-fit points are consistent with an inflation driven by a potential $V(\phi) \propto \phi^{2/5}$, whereas the other models require a more complicated functional shape.

A non-minimally coupled chaotic inflation model can be implemented by incorporating a term $(1 + \phi)R$ in the Lagrangian \cite{kallosh2025simple}. An extension of this framework is given by $(1 + \xi \phi)R$, which yields the predictions $n_s \simeq 1 - 3/2N_e$ and $r \simeq 4/\xi N_e^{3/2}$ \cite{kallosh2025present}. Using the best fit values of $r$ from table~\ref{tab:best} for all models, we obtain $\xi \sim (0.3,\, 0.6)$. These fall within the constraints from Planck-ACT-LB, which require $0.3 < \xi < 4$ \cite{ade2024constraining}.

\begin{table}[tbp]
\centering
\caption{Model diagnostics}
\label{tab:model_diag}
\begin{tabular}{lcc}
\toprule
Model & $\chi^{2}_{n_s}$ & $GT_{n_{s}}$ \\
\midrule
BH-1 & 3.147 & $0.13\sigma$ \\
BH-2 & 0.000 & $0.12\sigma$ \\
\midrule
T-1 & 1.044 & $0.29\sigma$ \\
T-2 & 0.001 & $0.32\sigma$ \\
\midrule
R-1 & 0.417 & $0.20\sigma$ \\
R-2 & 0.000 & $0.26\sigma$ \\
\midrule
K-1 & 1.831 & $0.20\sigma$ \\
K-2 & 1.036 & $0.12\sigma$ \\
\bottomrule
\end{tabular}
\end{table}

At the conclusion of the inflationary epoch, the energy density of the scalar field $\rho_\phi$ undergoes a transition, ultimately reheating the universe and establishing the initial conditions for the hot Big Bang \cite{cook2015reheating,german2024inflationary}. The dynamical evolution during this reheating phase can be characterized by the relationship between the energy density at the end of inflation ($\rho_e$) and that at the end of reheating ($\rho_{reh}$), governed by the equation of state parameter $w_{reh}$ during this phase \cite{munoz2015equation}:
\begin{equation}
\frac{\rho_e}{\rho_{reh}} = \left( \frac{a_e}{a_{reh}} \right)^{-3(1 + w_{reh})},
\label{eq:energy_ratio}
\end{equation}
where $a_e$ and $a_{reh}$ denote the scale factors at the end of inflation and at the completion of reheating, respectively.

The number of e-folds elapsed during the reheating epoch is defined as:
\begin{equation}
N_{reh} \equiv \ln\left( \frac{a_{reh}}{a_e} \right).
\label{eq:N_reh_def}
\end{equation}
The energy density at the end of inflation is determined by the scalar field potential $\rho_e \equiv \rho_\phi(\phi_e)$, where $\phi_e$ represents the field value at the termination of inflation.

Assuming instantaneous thermalization at the conclusion of reheating, the energy density $\rho_{reh}$ is related to the reheating temperature $T_{reh}$ via the standard radiation-dominated expression:
\begin{equation}
\rho_{reh} = \frac{\pi^2}{30} g_{reh} T_{reh}^4,
\label{eq:rho_rad}
\end{equation}
where $g_{reh} = 106.75$ corresponds to the effective number of relativistic degrees of freedom for the Standard Model plasma at high temperatures \cite{german2024inflationary}. Combining eqs.~\eqref{eq:energy_ratio}, \eqref{eq:N_reh_def} and \eqref{eq:rho_rad}, we derive a concise expression for the reheating temperature (in Planck units):
\begin{equation}
\ln(T_{reh}) = -\frac{3}{4}(1 + w_{reh}) N_{reh} + \frac{1}{4} \ln(\rho_e) - 0.890.
\label{eq:T_reh_final}
\end{equation}
Eq.~\eqref{eq:T_reh_final} establishes a direct relationship between the reheating temperature and the fundamental inflationary parameters $\rho_e$, $w_{reh}$ and $N_{reh}$, providing a crucial link between inflationary model predictions and observable cosmological quantities \cite{zharov2025reheating}.

\begin{table}[tbp]
\centering
\caption{Reheating parameters for different entropy models with varying equation of state parameter $w_{\text{reh}}$ and e-folding number $N_{\text{reh}}$.}
\label{tab:reheating_params}
\scriptsize
\begin{tabular}{lcccccccccccccc}
\toprule
Model & $\phi_e$ & $\rho_e \times 10^{-12}$ & \multicolumn{9}{c}{$\log_{10}(T_{\text{reh}}/\text{GeV})$} \\
\cmidrule(lr){4-12}
& & & \multicolumn{3}{c}{$w_{\text{reh}}=-1/3$} & \multicolumn{3}{c}{$w_{\text{reh}}=0$} & \multicolumn{3}{c}{$w_{\text{reh}}=1/3$} \\
\cmidrule(lr){4-6} \cmidrule(lr){7-9} \cmidrule(lr){10-12}
& & & $N_{\text{reh}}=10$ & $20$ & $30$ & $10$ & $20$ & $30$ & $10$ & $20$ & $30$ \\
\midrule
BH-1 & 0.121 & 6.839 & 13.036 & 10.865 & 8.693 & 11.951 & 8.693 & 5.436 & 10.865 & 6.522 & 2.179 \\
BH-2 & 0.116 & 5.934 & 13.021 & 10.849 & 8.678 & 11.935 & 8.678 & 5.421 & 10.849 & 6.507 & 2.164 \\
\midrule
T-1 & 0.120 & 5.169 & 13.006 & 10.834 & 8.663 & 11.920 & 8.663 & 5.406 & 10.834 & 6.491 & 2.149 \\
T-2 & 0.117 & 4.948 & 13.001 & 10.830 & 8.658 & 11.916 & 8.658 & 5.401 & 10.830 & 6.487 & 2.144 \\
\midrule
R-1 & 0.118 & 6.468 & 13.030 & 10.858 & 8.687 & 11.945 & 8.687 & 5.430 & 10.858 & 6.516 & 2.173 \\
R-2 & 0.112 & 6.211 & 13.026 & 10.854 & 8.683 & 11.940 & 8.683 & 5.426 & 10.854 & 6.511 & 2.169 \\
\midrule
K-1 & 0.121 & 6.256 & 13.027 & 10.855 & 8.684 & 11.941 & 8.684 & 5.427 & 10.855 & 6.512 & 2.169 \\
K-2 & 0.129 & 5.891 & 13.020 & 10.849 & 8.677 & 11.934 & 8.677 & 5.420 & 10.849 & 6.506 & 2.163 \\
\bottomrule
\end{tabular}
\end{table}

The sensitivity of the reheating temperature to variations in the input parameters can be quantified by considering fractional errors in eq.~\eqref{eq:T_reh_final}. A direct differentiating yields:

\begin{equation}
\frac{\Delta T_{reh}}{T_{reh}} = - \frac{3}{4}(1 + w_{reh}) \Delta N_{reh} + \frac{1}{4} \frac{\Delta \rho_e}{\rho_e}.
\label{eq:delta_T}
\end{equation}

This expression highlights two principal sources of uncertainty in determining $T_{reh}$, including variations in the duration of the reheating epoch ($\Delta N_{reh}$) and uncertainties in the inflationary energy scale ($\Delta \rho_e / \rho_e$). The prefactor $3(1+w_{reh})/4$ indicates that models with stiffer equations of state ($w_{reh} > -1/3$) exhibit stronger dependence on the reheating duration, while all models are equally sensitive to the precision of the inflationary energy density determination.

\begin{figure}[tbp]
\centering
\includegraphics[width=0.96\linewidth]{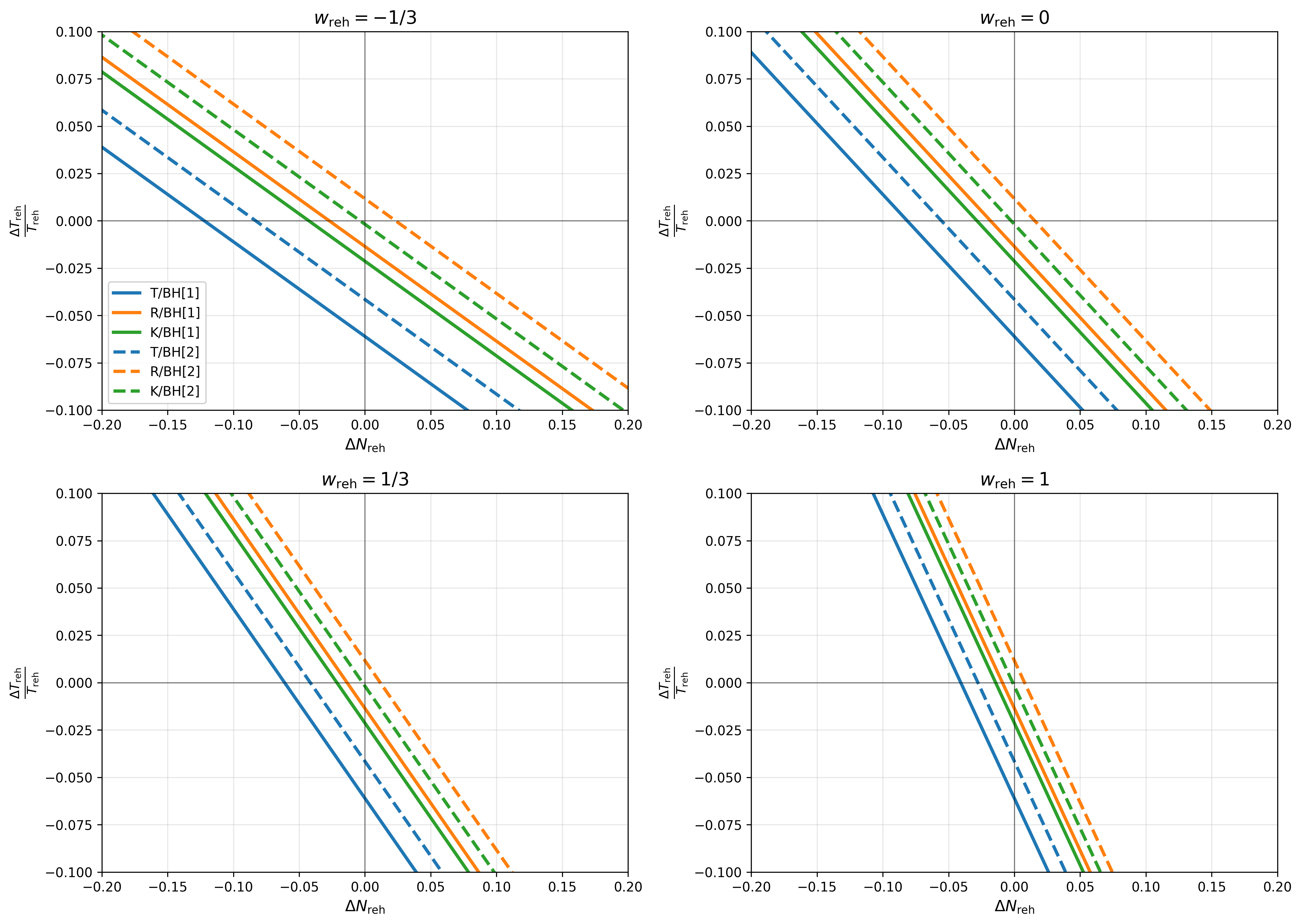}
\caption{Fractional error in temperature at end of reheating $\Delta T_{reh}/T_{reh}$, versus error in number of e-folds $\Delta N_{reh}$ during reheating epoch for T,R and K models relative to reference BH model. The plots are drawn in four panels correspond to various values of reheating effective EoS $w_{reh}$.}
\label{fig:Reheating}
\end{figure}

Our analysis of structure formation within the context of entropy-based inflationary models reveals several key insights into the connection between early-universe physics and late-time cosmic evolution. Utilizing the late universe as a $\Lambda$CDM framework, which is a valid leading-order approximation for our $f(\rho)$ function (see appendix~\ref{app:potential}), we have established a pipeline from fundamental inflationary parameters to observable quantities.

The analysis demonstrates that all considered entropy models produce values for the RMS mass fluctuation $\sigma_8$ in the range of $0.814$--$0.816$, showing remarkable consistency with current observational constraints from DESI and other surveys \cite{adame2025desi, mirpoorian2025dynamical}. The Gaussian Tensions ($GT_{\sigma_8}$) for all models remain below $1\sigma$, indicating good agreement with standard predictions. This similarity suggests that late-time structure formation is relatively insensitive to the specific entropy modification, provided the primordial power spectrum parameters---particularly the scalar spectral index $n_s$---are appropriately adjusted.

We further investigated the dependence of growth parameters on $n_s$. While the variations are small, they follow the expected trend where higher $n_s$ values correspond to slightly enhanced structure formation (see table~\ref{tab:cosmo_params_z0} in appendix~\ref{app:structure_calc}). The evolution of the key observable $f\sigma_8(z)$, which combines the growth rate and $\sigma_8$, shows the characteristic shape expected in $\Lambda$CDM while exhibiting minor amplitude variations across models (figure~\ref{fig:matterpower-fsigma8}).

The fractional differences in the matter power spectrum (right panels of figure~\ref{fig:matterpower-fsigma8}) highlight the scale-dependent impact of different primordial spectral tilts. These results underscore the sensitivity of cosmological observables to the primordial fluctuations generated during inflation.

In conclusion, the investigation of structure formation within generalized entropy frameworks demonstrates that late-time cosmological observables ($\sigma_8$, $f\sigma_8$, $S_8$) show minimal dependence on the specific entropy formulation when primordial parameters are properly constrained. These findings reinforce the viability of generalized entropy approaches to inflation while highlighting the critical importance of combining precise early and late-universe observations for comprehensive model testing.

\section{Conclusions}

In this work, we have investigated inflationary models within a systematic framework of generalized entropies by employing the Hamilton--Jacobi formalism with a non-linear parametrization of the Hubble parameter, $H(\phi) \sim 1 + h\phi^n$. By defining suitable likelihoods for the scalar spectral index ($n_s$) and the tensor-to-scalar ratio ($r$), and applying the NUT sampling algorithm, we constrained the free parameter spaces of these models. The formalism of~\cite{KHODAMMOHAMMADI2023138066} was applied, which embeds all entropic implications into a density function $f(\rho)$, whose deviation from the pivotal Bekenstein--Hawking (BH) model (with $f=1$) is fairly small. Overall, the Tsallis (T) entropy model exhibits the largest deviation from the standard BH case, while the Kaniadakis (K) entropy model shows the closest agreement. The models examined include BH, T, R\'enyi (R), and K entropies. The constraints on the free parameters were obtained as $\delta \simeq 1.1$--$1.2$ for T, $\alpha \sim \mathcal{O}(10^{-14})$ for R, and $K \sim \mathcal{O}(10^{-17})$ for K. We also derived a range for the number of e-folds, $N_e \sim 40$--$55$, consistent with conventional expectations for effective inflation.

We further studied the implications of these models for post-inflationary phases, including reheating and structure formation. In all cases, the differences between the models are minimal in this context, as expected, particularly for late-time structure formation, where variations in inflationary parameters play only a secondary role. Nevertheless, we highlighted subtle distinctions, such as the fractional sensitivity of the temperature relevant for BBN and small deviations from the pivotal curve in the matter power spectrum, in order to provide a comprehensive analysis.

Additionally, we examined the impact of varying the observational upper bound on $r$ and found that a slightly higher limit is marginally more compatible with current data within our framework. A lower bound of approximately $r \sim 0.01$ was obtained across all models, consistent with predictions of large-field inflationary scenarios with $\phi \sim 1$--$10\,M_{\mathrm{pl}}$. Moreover, we found that the desired non-linear parametrization of $H(\phi)$ implies $V(\phi)\propto \phi^{0.6}$, which is consistent with recent results from the ACT.

\section*{Acknowledgments}

The authors sincerely appreciate Supriya Pan for his constructive comments and insightful suggestions on this work.

\appendix
\section{Potential-Dominated Energy Density}
\label{app:potential}

The perturbed function of $f$ can be given by:
\begin{equation}
f = \sum_{i=0}^{N} f^{(i)},
\label{eq:perturbation_expansion}
\end{equation}
where $f^{(i)}$ is the $i$-th order of perturbation. Considering the entropy formulations in table~\ref{tab:entropies}, one can yield:
\begin{equation}
\begin{cases}
f^{(i)} \propto \delta_{i,0} & \text{for BH} \\
f^{(i)} \propto (\delta-1)^i & \text{for T} \\
f^{(i)} \propto (\alpha/H^2)^i & \text{for R} \\
f^{(i)} \propto (K/H^2)^{2i} & \text{for K}
\end{cases}
\label{eq:perturbation_orders}
\end{equation}
where $\delta_{i0}$ is the Kronecker delta. Therefore, according to best fits obtained for free parameters (table~\ref{tab:best}), one can neglect the $f^{(i)}$ for $i>1$. Also it's easy to see that $f^{(0)} = 1$ for all models, so we take $f \simeq 1 + f^{(1)}$ here. Looking to eq.~\eqref{eq:potential}, we obtain the factor of potential as $1/f^2 \simeq 1 - 2f^{(1)}$. By substituting perturbed $f$ into $V(\phi)$, and keeping up to first order of perturbation, one reads:
\begin{equation}
V(\phi) \simeq \frac{1}{8\pi^2} \left( 3\pi(1 - f^{(1)}) H^2 + (1 - 2f^{(1)}) \left( \frac{H_{,\phi}}{2} \right)^2 \right),
\label{eq:potential_perturbed}
\end{equation}
The first term is larger than second one by order of magnitude, so it's given by:
\begin{equation}
V(\phi) \simeq \frac{3}{8\pi} (1 - f^{(1)}) H^2,
\label{eq:potential_dominant}
\end{equation}
Taking the relation $3H^2/8\pi G = \rho(1 + f^{(1)})$, into account from first Friedmann eq.~\eqref{eq:Friedmann1}, we reach the desired equation as:
\begin{equation}
V(\phi) \simeq \rho(1 - (f^{(1)})^2) \simeq \rho,
\label{eq:potential_energy_equivalence}
\end{equation}
which demonstrates the potential-dominated nature of the energy density during slow-roll inflation in all entropy-modified scenarios.

The approximation in eq.~\eqref{eq:potential_energy_equivalence} holds under the slow-roll conditions where:
\begin{itemize}
\item Kinetic energy terms are subdominant: $\dot{\phi}^2/2 \ll V(\phi)$
\item Perturbative corrections remain small: $|f^{(1)}| \ll 1$
\item Hubble slow-roll parameters satisfy: $\epsilon_{mod}, \vert\eta_{mod}\vert \ll 1$
\end{itemize}

For our best-fit values of models, these conditions are well-satisfied, justifying the use of eq.~\eqref{eq:potential_approx} approximation in the main analysis.

\section{Validity of the Slow-roll Conditions}
\label{app:slow-roll}

In this appendix we derive the slow-roll parameters in terms of the Hubble parameter and its derivatives, and evaluate them for the corresponding model. The Hubble slow-roll parameters are defined as
\begin{align}
\epsilon_H &\equiv -\frac{\dot{H}}{H^2}, \\
\eta_H &\equiv -\frac{\ddot{\phi}}{H \dot{\phi}}.
\end{align}

Assuming $f \simeq 1$, and using the Hamilton--Jacobi relation
\begin{equation}
\dot{\phi} \simeq -\frac{1}{4\pi}\, H'(\phi),
\end{equation}
we can express these parameters in terms of derivatives of $H(\phi)$. From $\dot{H} = H'(\phi)\dot{\phi}$ and the above relation, one finds
\begin{align}
\epsilon_H &\simeq \frac{1}{4\pi} \left(\frac{H'(\phi)}{H(\phi)}\right)^2, \\
\eta_H &\simeq \frac{1}{4\pi} \frac{H''(\phi)}{H(\phi)}.
\end{align}
These expressions are consistent within the Hamilton-Jacobi formalism.

For the chosen functional form $H(\phi) = H_{\text{inf}}(h_0 + h_1 \phi^n)$,
\begin{align}
H'(\phi) &= H_{\text{inf}}\, h_1 n \phi^{n-1}, \\
H''(\phi) &= H_{\text{inf}}\, h_1 n (n-1)\phi^{n-2}.
\end{align}
Substituting into the slow-roll parameters yields
\begin{align}
\epsilon_H(\phi) &\simeq \frac{1}{4\pi}\,\frac{n^2 h_1^2 \phi^{2n-2}}{\left(h_0 + h_1 \phi^n\right)^2}, \\
\eta_H(\phi) &\simeq \frac{1}{4\pi}\,\frac{n(n-1) h_1 \phi^{n-2}}{h_0 + h_1 \phi^n}.
\end{align}

For the parameter values from table~\ref{tab:best}

\[
h_0 \simeq 0.4, \quad h_1 \simeq 0.6, \quad n \simeq 0.3, \quad \phi_0 \simeq 1.2,
\]

we obtain
\begin{align}
H(\phi_0) &\simeq 1.0341, \\
H'(\phi_0) &\simeq 0.1579, \\
H''(\phi_0) &\simeq -0.1019.
\end{align}
Thus,
\begin{align}
\epsilon_H(\phi_0) &\simeq 0.00186, \\
\eta_H(\phi_0) &\simeq -0.00783.
\end{align}

The slow-roll condition requires $\epsilon_H \ll 1$ and $|\eta_H| \ll 1$. At $\phi_0$, both conditions are satisfied:

\[
\epsilon_H \simeq 1.9 \times 10^{-3}, \quad |\eta_H| \simeq 7.8 \times 10^{-3}.
\]

Therefore, the model remains in the slow-roll regime at this field value, ensuring that inflationary expansion continues.

\section{Notes on Structure Formation}
\label{app:structure_calc}

This appendix details the calculations and numerical results related to structure formation discussed in section~\ref{sec:discussion}. We present the equations governing matter density evolution and linear perturbation growth, the definitions of key observables, and the specific results for our entropy-based inflationary models.

Assuming a flat universe, the evolution of the matter density parameter is given by:
\begin{equation}
\Omega_{m}(z) = \frac{\Omega_{m}(1 + z)^{3}}{\widetilde{H}^{2}(z)},
\label{eq:omega_m_evolution_app}
\end{equation}
where $\widetilde{H} \simeq \sqrt{\Omega_{m}(1 + z)^{3} + \Omega_{\Lambda}}$. The linear growth of matter perturbations is governed by:
\begin{equation}
\ddot{D} + 2H\dot{D} - 4\pi\rho_{m}D = 0,
\label{eq:growth_equation_app}
\end{equation}
where $D(z) = \delta_{m}(z)/\delta_{m}(0)$ is the linear growth factor. A solution can be written as:
\begin{equation}
D(z) = \exp\left( - \int_{0}^{z} \frac{f(z')}{1 + z'} dz' \right),
\label{eq:growth_solution_app}
\end{equation}
with the growth rate $f = d\ln D/d\ln a$ approximated by \cite{gong2009growth}:
\begin{equation}
f(z) = \Omega_{m}^{\gamma}(z),
\label{eq:growth_rate_app}
\end{equation}
and $\gamma \approx 0.55$ being the growth index \cite{linder2007parameterized}.

A key diagnostic is the RMS mass fluctuation in spheres of radius $8$ Mpc/$h$:
\begin{equation}
\sigma_{8}^{2} = \frac{1}{2\pi^{2}} \int_{0}^{\infty} P(k) W^{2}(kR) k^{2} dk,
\label{eq:sigma8_app}
\end{equation}
where $P(k)$ is the matter power spectrum and $W(kR)$ is the Fourier transform of a spherical top-hat filter. The power spectrum is modeled as:
\begin{equation}
P(k,z) = P_{\text{pri}}(k) T^{2}(k) D^{2}(z),
\label{eq:power_spectrum_app}
\end{equation}
with the primordial spectrum $P_{\text{pri}}(k) = A_{s} (k/k_{0})^{n_{s} - 1}$, where $k_{0} = 0.05$ Mpc$^{-1}$, and $T(k)$ is the transfer function.

We also compute the redshift-space distortion observable $f\sigma_{8}(z) = f(z)\sigma_{8}(z)$ and the weak lensing parameter $S_{8} = \sigma_{8}\sqrt{\Omega_{m}/0.3}$.

For our analysis, we utilized the $\mathtt{CAMB}$ code \cite{lewis2000efficient} to evaluate the matter power spectrum, incorporating transfer functions from Einstein-Boltzmann solutions and non-linear corrections via HaloFit \cite{takahashi2012revising}. We adopted Planck 2018 parameters: $H_{0} = 67.4$ km/s/Mpc, $\Omega_{b}h^{2} = 0.02237$, $\Omega_{c}h^{2} = 0.1200$, $\tau = 0.0544$, and $\Sigma m_{\nu} = 0.06$ eV \cite{miville2020planck}.

\begin{figure}[tbp]
\centering
\includegraphics[width=0.96\linewidth]{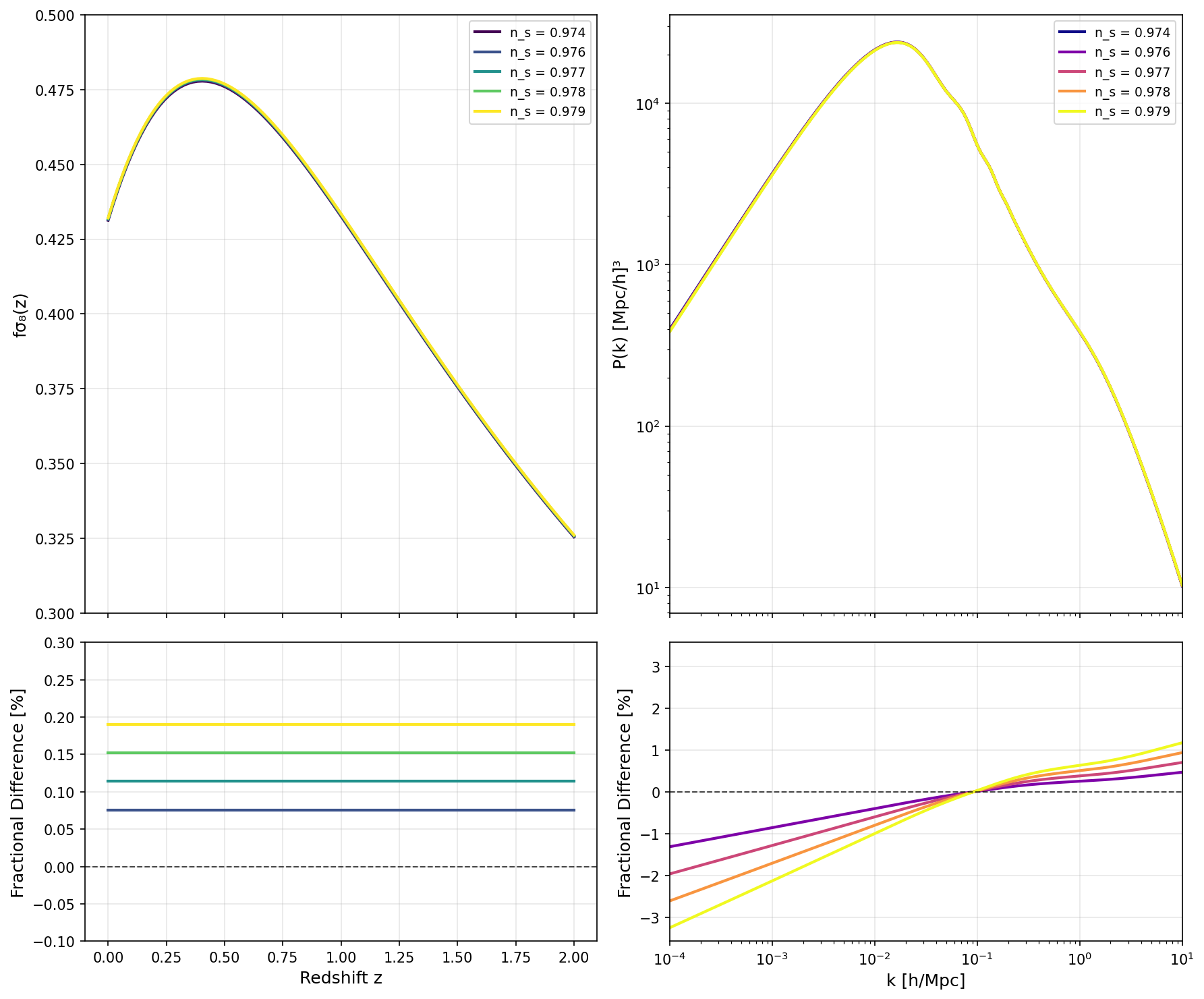}
\caption{Left: Evolution of $f\sigma_8(z)$ for selected primordial spectral indices $n_s$. Right: Matter power spectrum $P(k)$ at $z=0$ (top) and its fractional difference relative to the $n_s=0.974$ case (bottom), highlighting sensitivity to variations in $n_s$. The $n_s$ values correspond to best-fit models.}
\label{fig:matterpower-fsigma8}
\end{figure}

Table~\ref{tab:cosmo_params_z0} shows the growth parameters at $z=0$ for different spectral indices $n_s$ corresponding to the best-fit models. Figure~\ref{fig:matterpower-fsigma8} illustrates the matter power spectrum and $f\sigma_8$ evolution for selected $n_s$ values.

\begin{table}[tbp]
\centering
\caption{Growth parameters at $z=0$ for different spectral indices $n_s$ obtained from the best-fit model values.}
\label{tab:cosmo_params_z0}
\begin{tabular}{ccccc}
\toprule
\(n_s\) & \(\sigma_8(0)\) & \(f(0)\) & \(f\sigma_8(0)\) & \(S_8(0)\) \\
\midrule
0.974 & 0.814 & 0.530 & 0.431 & 0.834 \\
0.976 & 0.815 & 0.530 & 0.432 & 0.835 \\
0.977 & 0.815 & 0.530 & 0.432 & 0.835 \\
0.978 & 0.815 & 0.530 & 0.432 & 0.835 \\
0.979 & 0.816 & 0.530 & 0.432 & 0.836 \\
\bottomrule
\end{tabular}
\end{table}

\end{document}